\documentclass[12pt,preprint]{emulateapj}

\usepackage{graphicx}
\usepackage{amsmath}
\usepackage{hyperref}
\usepackage{color}

\hypersetup{
    colorlinks=true,
    linkcolor=red,
    citecolor=blue,
    urlcolor=magenta,
}

\begin{document}

\title{Constraints on compact dark matter with fast radio burst observations}
\author{Kai Liao$^{1}$}\email{liaokai@whut.edu.cn}
\author{S.-B. Zhang$^{2,3,4,5}$}
\author{Zhengxiang Li$^{6}$}
\author{He Gao$^{6}$}

\affil{
$^1$ {School of Science, Wuhan University of Technology, Wuhan 430070, China}\\
$^2${Purple Mountain Observatory, Chinese Academy of Sciences, Nanjing 210023, China}\\
$^3${University of Chinese Academy of Sciences, Beijing 100049, China}\\
$^4${CSIRO Astronomy and Space Science, PO Box 76, Epping, NSW 1710, Australia}\\
$^5${International Centre for Radio Astronomy Research, University of Western Australia, Crawley, WA 6009, Australia}\\
$^6$ {Department of Astronomy, Beijing Normal University, Beijing 100875, China}
}

\begin{abstract}
Fast Radio Bursts (FRBs) are bright radio transients with millisecond duration at cosmological distances. Since compact dark matter/objects (COs) could act as lenses and cause split of this kind of very short duration signals, Mu$\rm{\tilde{n}}$oz et al. (2016) has proposed a novel method to probe COs with lensing of FRBs. In this Letter, we for the first time apply this method to real data and give constraints of the nature of COs with currently available FRB observations.
We emphasize the information from dynamic spectra of FRBs is quite necessary for identifying any lensed signals and find no echoes in the existing data.
The null search gives a constraint comparable to that from galactic wide binaries, though the methods of redshift inference from dispersion measure would impact a little.
Furthermore, we make an improved forecast basing on the distributions of real data for the ongoing and upcoming telescopes.
Finally, we discuss the situation where one or more lensed signals will be detected. In such a case, the parameter space of COs can be pinned down very well
since the lens mass can be directly determined through the observed flux ratio and time delay between split images.

\end{abstract}
\keywords{fast radio bursts - gravitational lensing: strong - dark matter}

\section{Introduction}
A wide range of galactic and cosmological observations has verified the existing of dark matter, which contributes
a considerable part of the total energy density in the Universe. The $\textit{cold dark matter}$ model has successfully explained the observed large-scale structure.
However, we still know little about the constituent of dark matter on smaller scales and some issues exist in this standard model.
For example, according to the simulation, galaxies like the Milky Way should have
thousands of dark matter sub-halos surviving from the tide stripping process and appearing in the form of satellite dwarf galaxies, whereas only
$\sim 10$ such dwarfs have been observed in our galaxy and Andromeda M31 galaxy~\citep{Drlica-Wagner2015}.
Furthermore, one may conjecture that dark matter (or part of it) consists of
compact objects (COs), such as the massive compact halo objects (MACHOs)~\citep{Wyrzykowski2011,Pooley2009,Mediavilla2009,Monroy-Rodriguez2014}, primordial black holes (PBHs)~\citep{Carr1974,Carr1975},
axion mini-clusters~\citep{Hardy2017} and compact mini halos~\citep{Ricotti2009}.
For convenience, hereafter we take all of them as the compact dark matter/objects (COs). Some theoretical analysis allows the mass of COs to be as light as
$10^{-7}M_\odot$ and as heavy as the first stars $\sim 10^3M_\odot$~\citep{Griest1991}.

Probing COs through astronomical observation is therefore crucial to discriminate models and deepen our understandings on the nature of dark matter.
Efforts have been devoted with various approaches and some progress has been made in constraining the CO fraction in dark matter $f_{\rm{CO}}$ and the mass $M_{\rm{CO}}$.
While large-mass ($\geq 100M_\odot$) COs can
perturb the wide stellar binaries~\citep{Quinn2009}, the microlensing of stars can constrain the COs in the Milky way with
low-mass ($\leq10M_\odot$)~\citep{Tisserand2007,Wyrzykowski2011,Udalski2015,Calchi Novati2013,Niikura2017}.
Besides, by observing the lack of radiation as a result of accretion, one could also give a constraint for large-mass COs with the cosmic microwave background~\citep{Ali-Haimoud2017}.
Other methods include millilensing of quasars~\citep{Wilkinson2001}, lensing of supernovae~\citep{Benton2007}, ultra-faint dwarf galaxies~\citep{Brandt2016} and caustic crossing~\citep{Oguri2018}.
Generally speaking, no robust evidence of COs has been found for $f_{\rm{CO}}>0.1$ in a wide mass range.

The mass range $10-100M_\odot$ has been poorly constrained and
attracting most of the attention especially after the gravitational waves (GWs) from binary black holes were directly detected
by LIGO/VIRGO~\citep{Abbott2016}. The black hole masses are within such window, which suggests they could be the PBH dark matter~\citep{Bird2016,Sasaki2016}.
However, current constraints are too weak~\citep{Ricotti2008,Oguri2018}.
More robust and independent evidences are needed to verify such conjecture. Recently, lensing of transients like GWs~\citep{Jung2019,Liao2020},
gamma ray bursts (GRBs)~\citep{Ji2018} and fast radio bursts (FRBs)~\citep{Munoz2016} were proposed
to be very promising in constraining COs. The imprints of COs as lenses correspond to the distorted waveforms of GWs, the auto-correlation in GRB light curves and
the echoes of FRB signals.

Remarkably, FRB method should be the simplest and cleanest even though we have not understood the formation mechanism of FRB yet.
FRBs are bright pulses of emission at radio frequencies, most of which have durations of order milliseconds or less~\citep{Lorimer2007,Thornton2013}.
The short duration and large brightness make them emit coherently in nature. Most of FRBs seem to be one-off, but a few are repeaters manifesting a longer-lived central engine.
Recent studies showed it is possible that a large fraction (or even all) of the FRBs are repeaters, and we just happen to catch one of their bursts~\citep{Ravi2019,Munoz2020,Fonseca2020}.
While current event rate is limited by the small fields-of-view of current radio telescopes,
FRB events are supposed to be quite often on the full sky ($\sim10^4$/day).~\citep{Thornton2013,Champion2016}. The ongoing wide-field surveys like APERTIF, UTMOST, HIRAX and CHIME will
monitor a considerable fraction of the sky, giving thousands of detections per year. If part of dark matter consists of COs, there must be a chance that
an FRB is within the Einstein radius of a CO, appearing split signals with flux ratio and time delay. Therefore, detections of such lensed signals could statistically
infer the fraction and mass of COs in turn~\citep{Munoz2016}.
In principle, lensing of FRBs can effectively detect the mass range down to $20-100M_\odot$ that gives typical time delays comparable to the intrinsic duration of the signal.
Realistic constraints depend on the event number and distributions of signal durations and redshifts.
Shorter durations, higher redshifts and larger event number would give more stringent constraints.

The detected FRB events are timely included in the public catalogue\footnote{http://frbcat.org/}~\citep{Petroff2016}. The newest event number is $\sim110$ which gives a statistical sample.
We use these data to give a first constraint on COs and discuss more details about identifying the lensed signals in this work.
Besides, we also make corrections to the forecast and discuss how we will deal with the detected lensed FRBs.
The Letter is organized as follows: In Section 2, we introduce the theory on FRB lensing; In Section 3, we discuss how to identify the lensed signals
and apply our method to the existing data, giving the constraints; The forecast and lens mass estimation are shown in Section 4;
Finally, we summarize and make discussions in Section 5.

\section{Lensing of fast radio bursts}
Gravitational lensing is usually classified by the lens mass scale (equivalently the Einstein radius).
For FRB lensing, we suggest it be more appropriate to take it as strong lensing since we can clearly discriminate the split transient signals,
whereas the traditional microlensing limited by the resolution can only observe the overlapped images of constant sources. We take the CO as a point mass whose Einstein radius is given by
\begin{equation}
\theta_\mathrm{E}=2\sqrt{\frac{GM_\mathrm{CO}}{c^2D}}\approx(10^{-6})^{\prime\prime}\left(\frac{M_\mathrm{CO}}{M_\odot}\right)^{1/2}\left(\frac{D}{\mathrm{Gpc}}\right)^{-1/2},
\end{equation}
where the $\textit{effective lensing distance}$ (called $\textit{time delay distance}$ as well) $D=D_{\rm{L}}D_{\rm{S}}/D_{\rm{LS}}$, which is a combination
of three angular diameter distances. Subscripts $\rm{S,L}$ denote the source and the lens, respectively.
Although the spatial resolution in radio observation could reach very high level, for example, the angular resolution for the FRB 121102 with Very Long Baseline Array (VLBA) is $\sim(10^{-2})^{\prime\prime}$~\citep{Spitler2016,Chatterje2017,Tendulkar2017},
it is still insufficient to distinguish split images spatially for $M_{\rm{CO}}<10^8M_\odot$.
Therefore, we can not get the information of COs by measuring $\theta_{\rm{E}}$.
What one can directly measure is the time delay between the lensed signals, which is determined by
\begin{equation}
\Delta t=\frac{4GM_\mathrm{CO}}{c^3}(1+z_\mathrm{L})\left[\frac{y}{2}\sqrt{y^2+4}+\mathrm{ln}\left(\frac{\sqrt{y^2+4}+y}{\sqrt{y^2+4}-y}\right)\right],
\end{equation}\label{dt}
where the dimensionless impact parameter $y=\beta/\theta_E$ stands for the relative source position, $z_{\rm{L}}$ is the lens redshift.
Obviously, $\Delta t$ must be larger than the width ($w$) of the observed signal itself such that the split lensed images can be distinguished as double peaks.
This requires $y$ larger than certain value $y_{\rm{min}}(M_{\rm{CO}},z_{\rm{L}},w)$ according to Eq.\ref{dt}.

In addition, the flux/magnification ratio between two images ($+,-$) can be directly measured as well:
\begin{equation}
R_\mathrm{f}\equiv\left|\frac{\mu_+}{\mu_-}\right|=\frac{y^2+2+y\sqrt{y^2+4}}{y^2+2-y\sqrt{y^2+4}}>1.\label{Rf}
\end{equation}
To make both lensed images (especially the fainter one) detectable with high enough signal-to-noise ratio (SNR), $R_{\rm{f}}$ should not be too large,
which requires the impact parameter to be smaller than certain value $y_{\rm{max}}=\left[(1+R_{\rm{f,max}})/\sqrt{R_{\rm{f,max}}}-2\right]^{1/2}$.
We set the criterion $R_{\rm{f,max}}=5$ following Mu$\rm{\tilde{n}}$oz et al. (2016).

For a given FRB event at $z_{\rm{S}}$, the lensing optical depth is the probability that the point source is within the perceptible region of any COs along the
line of sight:
\begin{equation}
\tau(M_\mathrm{CO},f_\mathrm{CO},z_\mathrm{S},w)=\int_0^{z_\mathrm{S}}d\chi(z_\mathrm{L})(1+z_\mathrm{L})^2n_\mathrm{CO}\sigma(M_\mathrm{CO},z_\mathrm{L},z_\mathrm{S},w),\label{taun}
\end{equation}
where $\chi$ is the comoving distance, $n_{\rm{CO}}$ is the CO number density and the cross section is given by
\begin{equation}
\sigma(M_\mathrm{CO},z_\mathrm{L},z_\mathrm{S},w)=\frac{4\pi GM_\mathrm{CO}}{c^2}\frac{D_\mathrm{L}D_\mathrm{LS}}{D_\mathrm{S}}\left[y_\mathrm{max}^2-y_\mathrm{min}^2(M_\mathrm{CO},z_\mathrm{L},w)\right].
\end{equation}
Using Hubble parameter at lens redshift and Hubble constant, Eq.\ref{taun} can be rewritten as:
\begin{equation}
\begin{aligned}
\tau(M_{\rm{CO}},f_{\rm{CO}},z_S,w)=\frac{3}{2}f_{\rm{CO}}\Omega_c\int_0^{z_{\rm{S}}}dz_{\rm{L}}\frac{H_0^2}{H(z_{\rm{L}})}\frac{D_{\rm{L}}D_{\rm{LS}}}{D_{\rm{S}}}    \\
\times(1+z_{\rm{L}})^2\left[y_\mathrm{max}^2-y_\mathrm{min}^2(M_\mathrm{CO},z_\mathrm{L},w)\right].  \label{tau}
\end{aligned}
\end{equation}
We adopt the flat $\Lambda$CDM cosmology with total dark matter density $\Omega_c=0.24$, baryonic matter density $\Omega_b=0.06$ and
Hubble constant $H_0=70\rm{km\ s^{-1}Mpc^{-1}}$. Following other works, we assume a fraction of dark matter is in
the form of COs having the same mass $M_{\rm{CO}}$, then $f_{\rm{CO}}=\Omega_{\rm{CO}}/\Omega_c$.

According to the definition, the expected number of lensed FRBs is the sum of the lensing optical depths of all FRBs (for $\tau_i\ll1$):
\begin{equation}
N_\mathrm{lensed}(M_\mathrm{CO},f_\mathrm{CO})=\sum_{i=1}^{N_\mathrm{total}}\tau_\mathrm{i}(M_{\rm{CO}},f_{\rm{CO}},z_{S,i},w_i).\label{N}
\end{equation}
This equation shows that for the given ($M_\mathrm{CO},f_\mathrm{CO}$), it will predict the corresponding number of detectable lensed FRB signals. On the contrary,
one can infer ($M_\mathrm{CO},f_\mathrm{CO}$) with the number of observed lensed signals. Particularly, if no lensed signal is detected, the region
in ($M_\mathrm{CO},f_\mathrm{CO}$) parameter space that predicts at least one lensed signal should be ruled out, which is the standard analysis pipeline widely used in the literature.

\section{constraints with current observations}
The number of verified FRBs is rapidly increasing. At the moment of writing this Letter, the reported FRB number is 110.
In addition, there are extra 9 events that are highly considered as the candidates. Although the method only requires
the transient nature, most of the candidates do not have the measured widths of the signals and are therefore not used by us in this work.
We will introduce how we analyze these data and constrain COs in this section.

\subsection{Identifying the lensed signals}

\begin{figure}
\begin{center}
\begin{tabular}{cc}
\includegraphics[width=8cm,angle=0,trim={25mm 15mm 58mm 30mm}]{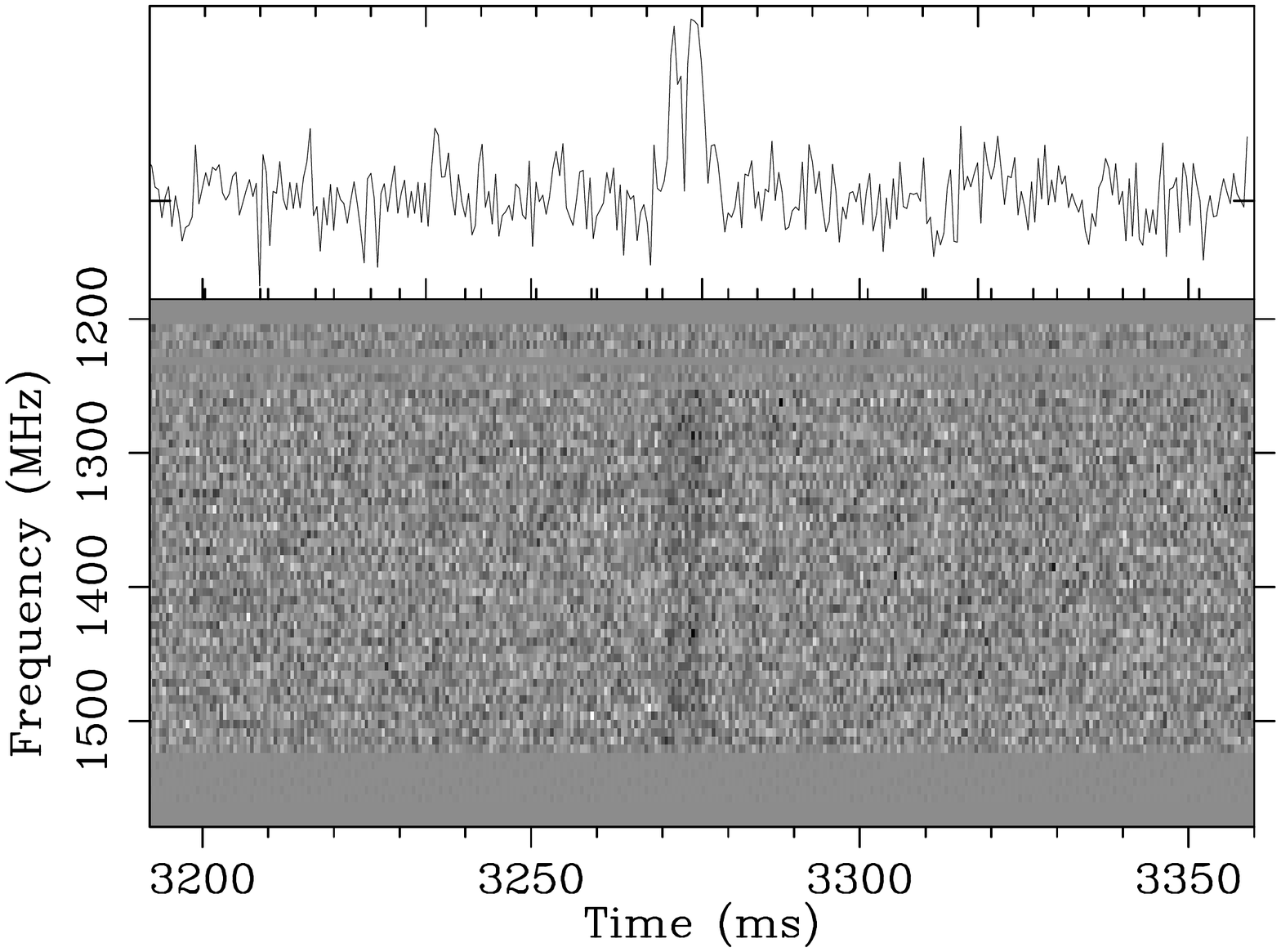} \\
\includegraphics[width=8cm,angle=0,trim={25mm 15mm 58mm 30mm}]{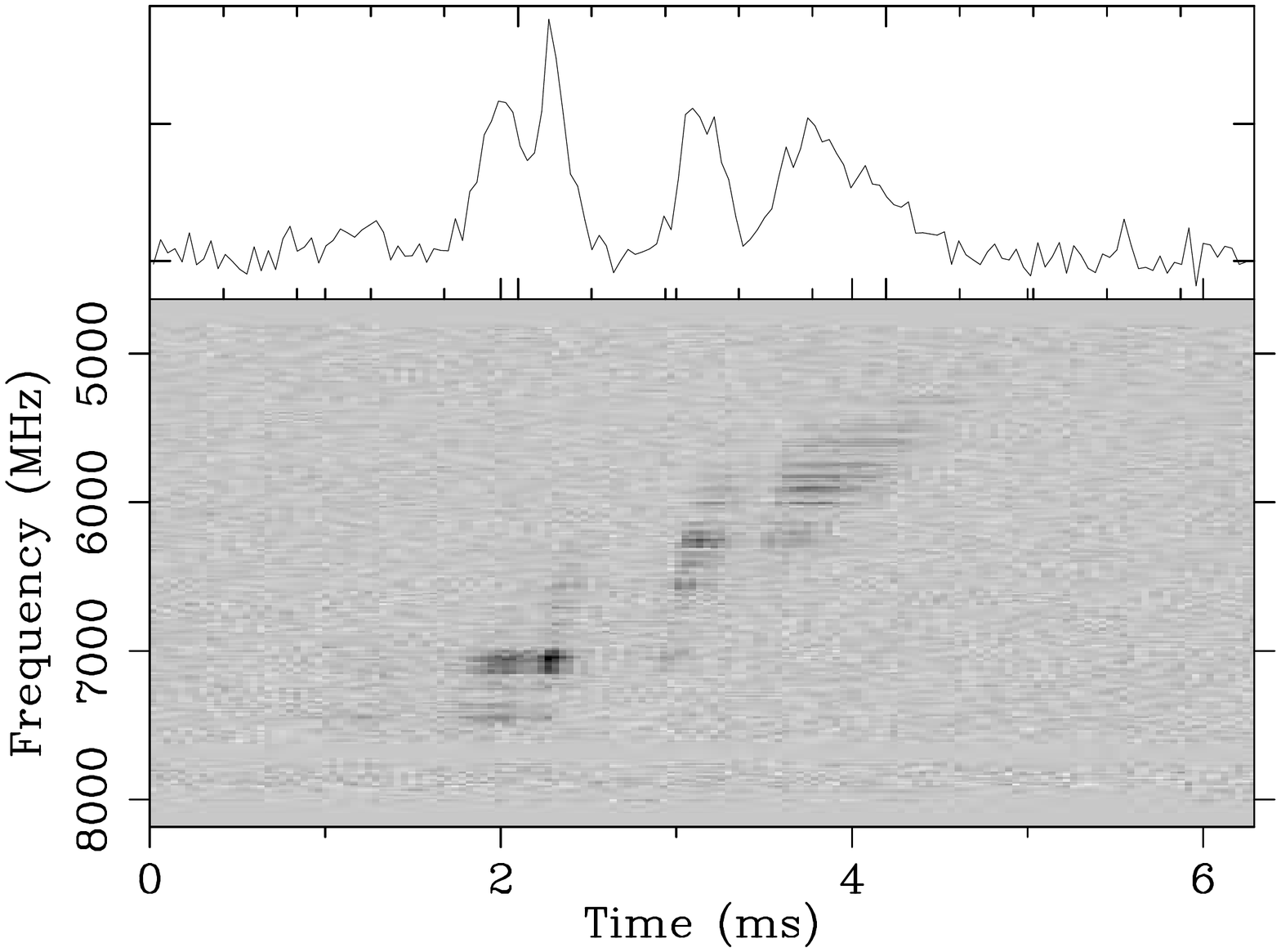}\\
\end{tabular}
\caption{Dynamic spectra of FRB~121002 and a multiple-peak burst of FRB~121102 plotted using the raw data from \protect\url{https://data-portal.hpc.swin.edu.au/dataset} and \protect\url{http://seti.berkeley.edu/frb121102}, respectively.
}
\label{figure:mp_FRBs}
\end{center}
\end{figure}

In Mu$\rm{\tilde{n}}$oz et al. (2016), the double-peak structure was pointed out to be the feature of a lensed FRB.
We have searched such signals in the catalog and find a few existing FRBs that have multiple-peak structure and are likely to be lensed. They are
FRB 170827~\citep{Farah2018}, FRB 121002~\citep{Champion2016}, FRB 121102 (repeating)~\citep{Hessels2018}, FRB 180814.J0422+73 (repeating)~\citep{CHIME/FRB2019}, FRB 181112~\citep{Cho2020} and the very recent FRB 181123 by the Five-hundred-meter Aperture Spherical radio Telescope (FAST)~\citep{Zhu2020}.
To identify a lensed signal, we suggest that one should further use the dynamic spectrum information which reflects the intrinsic feature of an FRB.
The dynamic spectra of these events are presented in the original papers except for FRB 121002. One can easily tell that FRB 170827, 121102, 180814.J0422+73 and 181123 are not lensed since the pulses in the dynamic spectra corresponding to different peaks show different structures.
It is impossible to fit them using a simple time delay and relative magnification parameters like what lensing requires.
Rather than lensing effect, the multiple peaks of these FRBs must come from the intrinsic substructure of the signals themselves.
For example, FRB 181123 has three peaks, but peak 2 only has the higher and peak 3 only has the lower frequency parts compared to peak 1.
We plot the dynamic spectrum of FRB 121002 in the upper panel of Fig.\ref{figure:mp_FRBs}. Lensing of this event was first discussed by~\citep{Munoz2016}. The signal-to-noise ratio is small such that the dynamic spectrum can not easily distinguish
the two peaks and we can not compare them. However, the second arrived peak has an intenser pulse than the first one, which is against the prediction of lensing theory. We therefore take it as an unlensed event. We also plot the repeating FRB 121102 in the lower panel for example.
It clearly shows the ``frequency drift" phenomenon where multiple bursts occur within several milliseconds with
decreasing frequencies.
At last, it is worth mentioning that the spectrum of FRB 181112 showed two similar pulses with very large flux ratio~\citep{Cho2020}.
However, the different polarization details and
the impossibility of wave effects indicate the peaks should be intrinsic~\citep{Cho2020}.

Therefore, we emphasize that
it is important to use more information like the dynamic spectra or polarization properties to identify any lensed signals such that the degeneracy between intrinsic substructure and lensing can be broken.
A lensed FRB should appear in dynamic spectrum as
two pulses with the same shape and only different from each other by flux magnification and time delay (the fainter one comes later as the echo).
We have carefully examined the dynamic spectra of the 110 FRBs with their original papers or the raw data on the FRB website, especially those who have multiple peaks. No strong evidence
of lensing signal was found, which can shed light on the properties of lenses.

\subsection{Results}
The radio pulse from FRB experienced a frequency-dependent delayed time through the ionized interstellar medium, quantified by a dispersion
measure (DM) which is proportional to the number of electrons along the line of sight. If we know the ionized history of the Universe, we
can infer the distances/redshifts with the directly measured DMs.
The observed DM of an FRB can be decomposed into
\begin{equation}
\mathrm{DM=DM_{MW}+DM_E},
\end{equation}
where
\begin{equation}
\mathrm{DM_E=DM_{IGM}+\frac{DM_{host}+DM_{src}}{1+z}}
\end{equation}
is the external DM contribution outside the Milky Way galaxy, and $\rm{DM_{host}}$ and
$\rm{DM_{src}}$ are from FRB host galaxy and source environment, respectively.
The biggest issue in this manner is we have not much information on the host galaxy and source environment, except for those who can be localized~\citep{Li2020}.
The current viewpoint is that the average $\rm{DM_{host}+DM_{src}}$ could span from several tens to $\sim200{\rm\ pc~cm^{-3}}$. To make a robust conclusion, we adopt
the maximum value $200{\rm\ pc~cm^{-3}}$, equivalently the minimum inference of $z$ for all host galaxies.
For the galactic DM contribution $\rm{DM}_{\rm{MW}}$, we use the NE2001 galactic electron density model~\citep{Cordes2002}.
We adopt two intergalactic medium (IGM) models of inferring redshifts from the rest of dispersion measure $\rm{DM_{IGM}}$.
In model 1, we follow the original work by \citet{Petroff2016}, where the fraction of baryon mass in the IGM $f_{\rm IGM}$ was supposed to be unity ($f_{\rm IGM}=1.0$) and the He ionization history was not taken into consideration, approximately ${\rm DM_{IGM}} \sim 1200z~{\rm pc~cm^{-3}}$~\citep{Ioka2003}.
In model 2, the ${\rm DM_{IGM}}-z$ relation is given by \citet{Deng2014},
approximately ${\rm DM_{IGM}} \sim 855z~{\rm pc~cm^{-3}}$~\citep{Zhang2018}, with the consideration of He ionization history and $f_{\rm IGM}=0.83$.
With the current five localized FRBs, model 2 seems to be more favored~\citep{Li2020}.

\begin{figure}
\includegraphics[width=\columnwidth,angle=0]{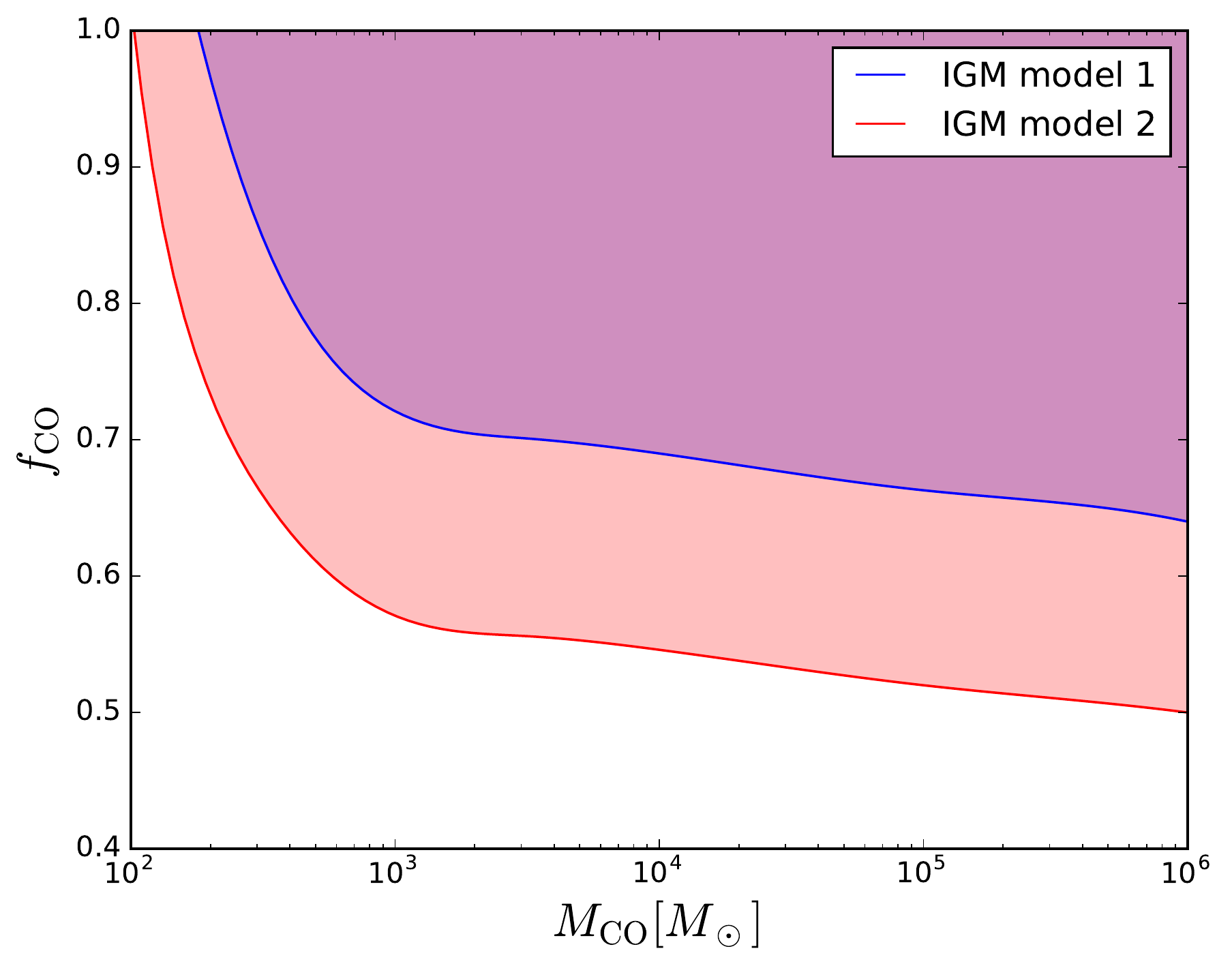}
   \caption{Constraints on the CO fraction and mass based on the fact that no lensed signal has been found in current data. The shaded regions are ruled out.
   The limits are at $68\%$ confidence level (1 $\sigma$) and there are no limits within 2 $\sigma$.
  }\label{results}
\end{figure}

We follow the standard operating procedure in the literature for studying the nature of COs.
For each ($M_{\rm{CO}},f_{\rm{CO}}$) point in Fig.\ref{results}, it corresponds to an
expected number of lensed FRB signals according to Eq.\ref{N}. Since no lensed signal has been
found in the current data, the shaded regions in the ($M_{\rm{CO}},f_{\rm{CO}}$) parameter space that predict
at least one detectable lensed signal should be ruled out at $68\%$ confidence level (1 $\sigma$).
In the case of IGM model 2, the mass can be tested down to $\sim 100M_\odot$ and $f_{\rm{CO}}$
is gradually constrained to $\sim0.5-0.6$ for large mass. While in the case of IGM model 2, the constraints are weaker since it gives smaller redshifts.
Our results are comparable to that from
wide binaries~\citep{Quinn2009}. Although current constraints are relatively weak, especial for small masses, we have proved the feasibility of this method.
For thousands of events detected in the near future, we will give a much better constraint, especially for small masses ($<100M_\odot$).

\section{forecast}
In this section, we use the realistic distributions of the data to make an improved forecast.
Furthermore, we discuss how COs can be constrained with detected lensed signals.

\begin{figure}
\includegraphics[width=\columnwidth,angle=0]{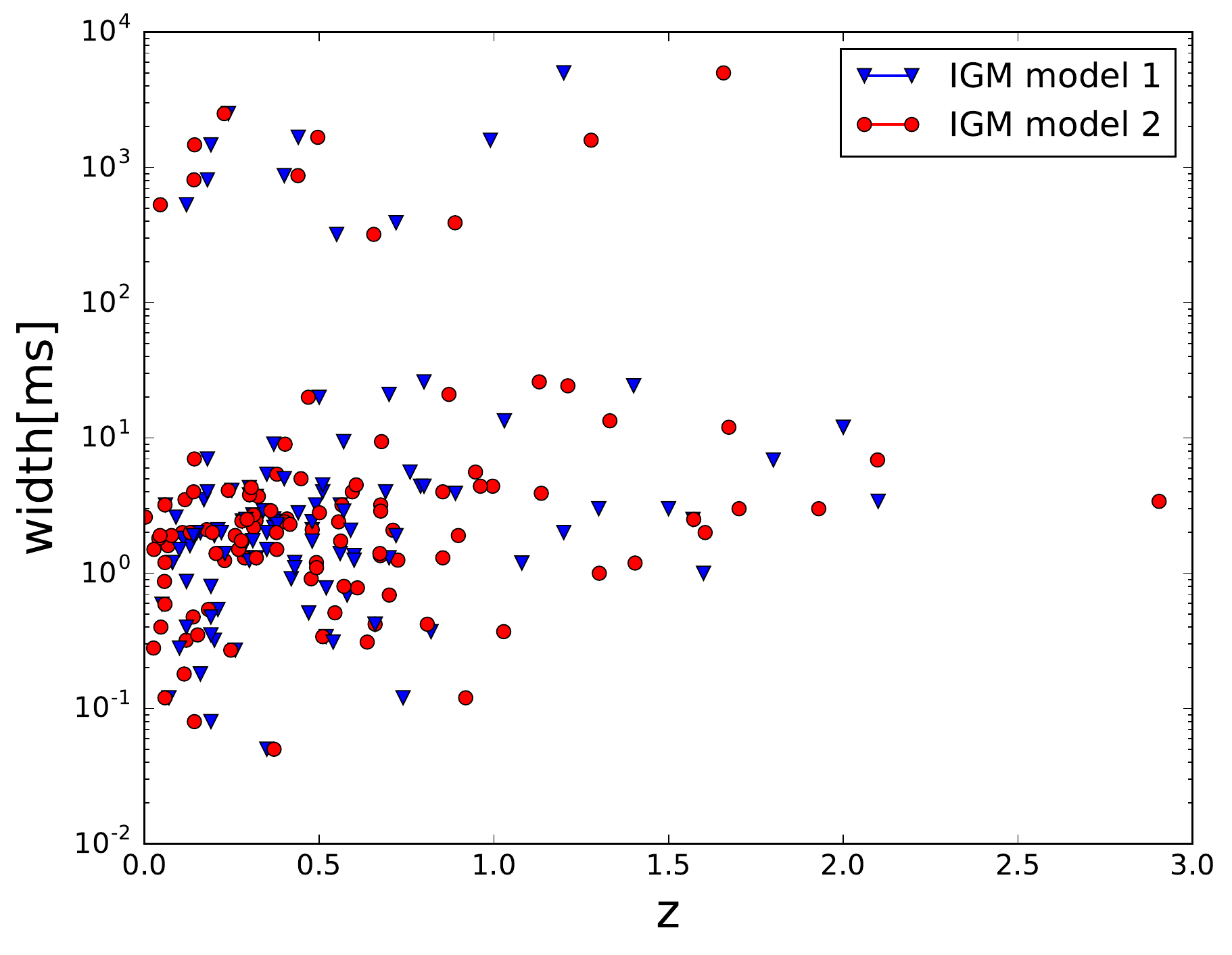}
   \caption{The 2-dimensional distribution of widths and inferred redshifts with two methods. We note that a width desert between 30 and 300 ms exists in current data.
  }\label{distribution}
\end{figure}

\begin{figure}
\includegraphics[width=\columnwidth,angle=0]{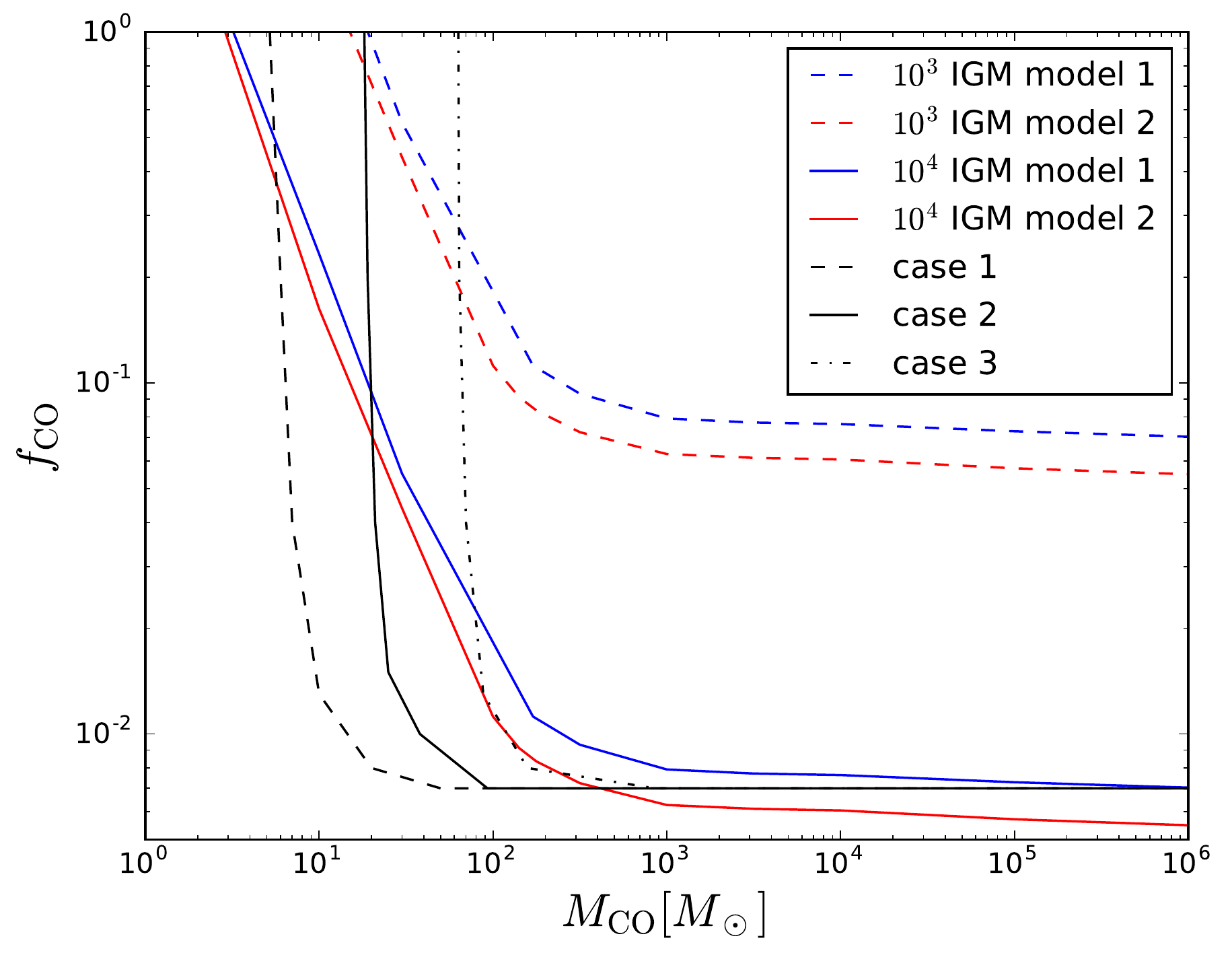}
   \caption{An forecast based on the realistic distributions of the data. The critical lines (red and blue) correspond to the cases that one lensed signal is expected to be detected.
   We also show the results by Mu$\rm{\tilde{n}}$oz et al. (2016) for comparison, where the event number is $10^4$, the constant signal widths are 0.3, 1 and 3 ms in case 1, 2 and 3, respectively.
  }\label{forecast}
\end{figure}

\subsection{A null search case}
In Mu$\rm{\tilde{n}}$oz et al. (2016), to calculate the integrated lensing probability, the optical depth for lensing of a
single burst had to be convolved with the redshift distribution of FRBs. They assumed FRBs either have a constant comoving number density
or a scenario where FRBs follow the star-formation history. Since we know little about the FRB origin and the DM contribution from host galaxies, there is no reason to make
any assumptions for redshift distribution of FRBs. For example, if the progenitors of FRBs are binary stars, then a delay time distribution (DTD) relative to the star formation rate exists.
The direct and more robust way is to understand FRB redshifts from the detected signals themselves.
Furthermore, they assumed a constant width of FRB to be 0.3, 1 and 3 ms, respectively, which is not realistic. We make forecast basing on the real distribution of the data.
The 2-dimensional distribution of widths and redshifts is plotted in Fig.\ref{distribution}. The widths are observed ones, rather than the intrinsic.
The data are from the FRB website http://frbcat.org/ which provides the observed (inferred as well) parameters of each verified signal. The redshifts are inferred
from the dispersion measure with two IGM models.
The generated events for forecasting follow the 2-D distribution in Fig.\ref{distribution}, i.e., the simulation follows the observed data themselves. The number $10^4$ we assume in this work do not rely on certain surveys. The
data from all the telescopes could be used in the analysis. The number is chosen such that we can compare our results with Mu$\rm{\tilde{n}}$oz et al. (2016).
Besides, $10^4$ FRBs per year is promising for CHIME-like telescopes.
The improved forecast is shown in Fig.\ref{forecast}.
The critical curves are similar to those in Mu$\rm{\tilde{n}}$oz et al. (2016), however, they are less steep for the small-mass end determined by some very small widths in the catalog, while the decreasing trend persists to large mass due to some very large widths.
In addition, we also consider $10^3$ events for either very near future or a pessimistic scenario.

\subsection{Constraints from lensed signals}
We discuss the case that at least one lensed signals will be verified.
Once a lensed FRB signal can be detected, we can estimate the lens mass from the measured time delay and flux ratio.
The source position can be determined from flux ratio, then the redshifted lens mass can be determined from time delay.
Compared to the uncertainties in the measured time delay and flux ratio, the uncertainty of lens redshift dominates.
The typical value is $\sigma_{z_L}\sim 0.5$. Nevertheless, it is sufficient for current CO studies. The mass can be pinned down very well on certain scale.
Moreover, if more than one lensed signals are detected, we can even test whether COs consists of the same mass and
the theories that give a non-constant mass function. The intermediate-mass black holes may also be discovered in this way.

\begin{figure}
\begin{center}
\begin{tabular}{cc}
\includegraphics[width=8cm,angle=0,trim={25mm 15mm 58mm 30mm}]{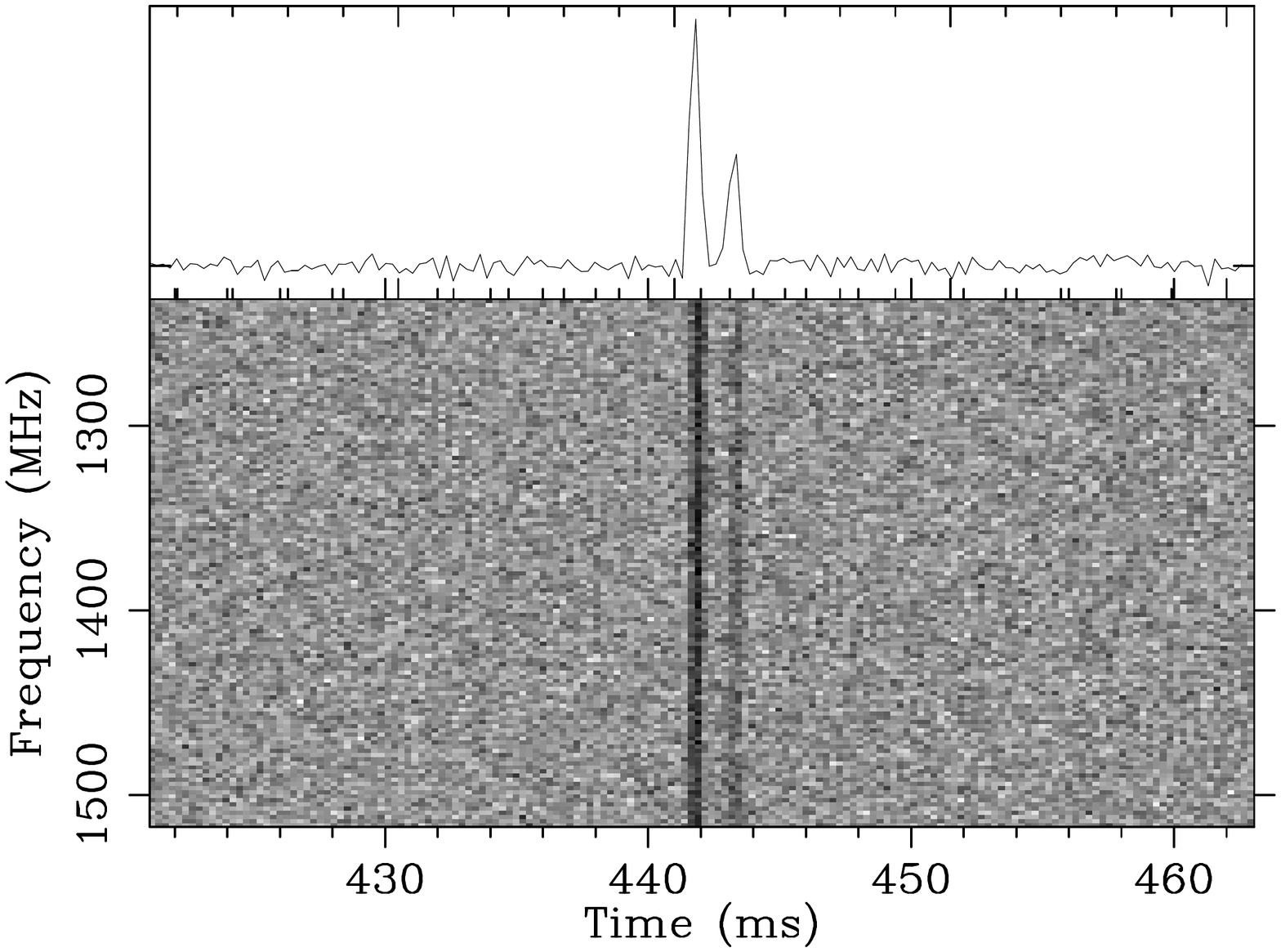}\\
\end{tabular}
\caption{Dynamic spectrum of a simulated lensed signal.}
\label{figure:sim_FRB}
\end{center}
\end{figure}

To show how this method works, we simulate a typical lensed signal in Fig.\ref{figure:sim_FRB} for example. It uses \emph{\sc psrfits} search mode format covering a frequency range of 1230$-$1518\,MHz of 512 channels. The data are two-bit sampled with sampling time of 64\,$\mu$s. The DMs and widths at 50 \% power point of these two pulses are 1000\,cm$^{-3}$\,pc and 0.5\,ms, respectively.
The redshift of the FRB $z_{\rm{S}}=1.0$ and the compact dark matter
is located at $z_{\rm{L}}=0.5$. The source position relative to the Einstein radius is $y=0.5$. Assuming a flat $\Lambda$CDM model with $\Omega_{\rm{M}}=0.3$ and $H_0=70\rm{km/s/Mpc}$, the time delay between two lensed signal can be got $\Delta t=1.5$ ms, the first arrived signal has magnification $|\mu_+|=1.6$ and the second signal has $|\mu_-|=0.6$, with flux ratio $R_{\rm{f}}=2.7$.
From the dynamic spectrum, one can clearly see two identical pulses except for the time delay and flux ratio differences. If we detect such a signal with $\Delta t$ and $R_f$ measurements,
we can infer the redshifted lens mass $M_z=M_{\rm{CO}}(1+z_{\rm{L}})=45M_\odot$ basing on Eq.\ref{dt} and Eq.\ref{Rf}. However, since $z_{\rm{L}}$ can not be observed, $M_{\rm{CO}}=M_z/(1+z_{\rm{L}})$ inference has
the uncertainty from $z_{\rm{L}}$ which should be in $[0,z_{\rm{S}}=1]$, giving $22.5M_\odot<M_{\rm{CO}}<45M_\odot$. Nevertheless, since we have information on $M_{\rm{CO}}$, the degeneracy between $M_{\rm{CO}}$ and $f_{\rm{CO}}$
can be broken. This would further narrow the allowed regions in Fig.\ref{results} or Fig.\ref{forecast}. This idea is very similar to the mass estimation in lensing of gravitational waves~\citep{Cao2014}.

\section{Summaries and prospectives}
Fast radio bursts are one of the most exciting new mysteries of astrophysics.
Beyond how they are created, there is also the prospect of using FRBs to probe the extremes of the Universe and
the invisible intervening medium. Due to the short duration, cosmological distances and the large event rate,
the lensing of FRBs could be a powerful and robust tool to probe the compact dark matter/objects.
We have made some progress in this work summarized as follows:

\begin{enumerate}
\item For the first time, we use realistic FRB data to give a constraint on the fraction and mass of COs.
      The constraint results are comparable to that from wide binaries;
\item We make an improved forecast basing on the distributions of the existing FRBs for the upcoming CHIME-like experiments;
\item We discuss the importance of using dynamic spectra of FRBs in identifying the lensed signals. It can effectively break the degeneracy between intrinsic structure and lensing imprints;
\item We discuss the situation when a few lensed signals can be detected and find the CO parameter space can be further well determined by lens mass estimation.
\end{enumerate}

For future studies, it is necessary to build up an effective pipeline to identify lensed FRBs, especially for the upcoming large number of FRBs.
It is also important to understand the properties of the host galaxies, the ionization history of the Universe and its fluctuation in each direction such that the redshift inference
can be more accurate. Fast and high spatial resolution program will directly find the host galaxies, thus a large number of redshifts can be measured accurately.
More and more events are having polarization measurements, which can be used as an extra criterion for identifying lensed signals, especially for those with similar dynamic spectra, since lensing would not change the polarization.
While we are writing this Letter, we note a very recent work based on analyzing FRB 181112 and 180924~\citep{Sammons2020}. It shows the burst substructure with high time resolution can be measured down to $15\mu s$ such that much smaller mass scales can be probed, making FRB method very promising.

\section*{Acknowledgments}
We thank the referee for his/her helpful comments and
G. Hobbs for his \emph{\sc pfits} software package to simulate the lensed signal.
KL was supported by the National Natural Science Foundation of China (NSFC) No.~11973034.
ZL was supported by the NSFC No.~11920101003. HG was supported by the NSFC No.~11722324, 11690024, 11633001, the Strategic Priority Research Program of the Chinese Academy of Sciences No.~XDB23040100 and the Fundamental Research Funds for the Central Universities. SBZ was supported by the NSFC No.~11725314.

\clearpage

\end{document}